# Ultrafast all-optical second harmonic wavefront shaping

A. Sinelnik[1,2†], S. H. Lam[2†], F. Coviello[1,2,3†], S. Klimmer[1], G. Della Valle[3,4], D.-Y.Choi[5], T. Pertsch[2,6,7], G. Soavi[1], I. Staude[1,2,7*]

[1]*Friedrich Schiller University Jena, Institute of Solid State Physics, 07743 Jena, Germany*

[2]*Institute of Applied Physics, Abbe Center of Photonics, Friedrich Schiller University Jena, 07745 Jena, Germany*

[3]*Dipartimento di Fisica, Politecnico di Milano, Piazza Leonardo da Vinci, 32, 20133 Milano, Italy*

[4]*Istituto di Fotonica e Nanotecnologie, Consiglio Nazionale delle Ricerche, Piazza Leonardo da Vinci, 32, 20133 Milano, Italy*

[5]*Laser Physics Centre, Research School of Physics, Australian National University, Canberra ACT 2601, Australia*

[6]*Fraunhofer Institute for Applied Optics and Precision Engineering, 07745 Jena, Germany*

[7]*Max Planck School of Photonics, 07745 Jena, Germany*

isabelle.staude@uni-jena.de

† These authors contributed equally to this work

*Corresponding author isabelle.staude@uni-jena.de

**Optical communication can be revolutionized by encoding data into the orbital angular momentum of light beams. However, state-of-the-art approaches for dynamic control of complex optical wavefronts are mainly based on liquid crystal spatial light modulators or miniaturized mirrors, which suffer from intrinsically slow (µs-ms) response times. Here, we experimentally realize a hybrid meta-optical system that enables complex control of the wavefront of light with pulse-duration limited dynamics. Specifically, by combining ultrafast polarization switching in a WSe2 monolayer with a dielectric metasurface, we demonstrate second harmonic beam deflection and structuring of orbital angular momentum on the femtosecond timescale. Our results pave the way to robust encoding of information for free space optical links, while reaching response times compatible with real-world telecom applications.**

Three decades ago, Allen et al. demonstrated that, in addition to linear momentum and spin angular momentum, photons propagating in paraxial vortex beams possess orbital angular momentum (OAM)[1,2], or topological charge, which takes values that are integer multiples of the reduced Planck constant. The OAM is associated with the global structure of light and the topological charges correspond to the number of intertwined helical surfaces in the wavefront of a light field[3-5]. Importantly, OAM light beams with different integer topological charge represent orthogonal states, even for finite apertures and regardless the aperture size[6]. As a consequence, they are exceptionally robust against perturbations, making them a viable platform for the encoding of digital information of both classical and quantum nature[7-10], and an interesting candidate to replace commonly used amplitude or combined amplitude/phase encoding schemes. In particular, OAM beams hold great potential for free space optical communications, where atmospheric perturbations pose an omnipresent challenge[7]. Moreover, it is impossible to recover the topological charge from the light scattered by the atmosphere, because the time dependent scattering processes randomize the phase structure of the beam[11]. Consequently, OAM encoded information is resistant to eavesdropping, adding the paramount feature of security to its potential for free-space optical communication.

While being highly promising, the main missing piece to the spread of free-space optical communication based on encoding of digital information into OAM is the lack of suitable methods for ultrafast wavefront shaping, which is required for the dynamic generation of topological charges, as opposed to the simpler modulation of intensity or temporal phase. As a consequence, the speed at which information can be encoded via the OAM is still limited. To date, the most common approach was based on liquid crystal (LC) spatial light modulators (SLMs)[12] with response times of tens of ms, restricting data rates to the kbit/s range[13]. To reduce the response times down to μs, technologies such as ferroelectric LC-SLMs[14], digital micromirror devices (DMDs)[15] and photothermal SLMs[16] have been suggested. DMDs in particular have already been used for OAM encoding with a record high modulation rate of 17.8 kHz[17], which is however still far from the demands of realistic telecom applications. For this reason, light structuring has been exploited for multiplexing, but not for the encoding of optical bits in telecom applications[18].

As such, the possibility to switch vortex beams at GHz rate remains a major task for the success of OAM based communication and information technology. More broadly, the capability to control the wavefront of light at ultrafast time scales would also enable disruptive developments in other technological areas, such as beam steering, quantum optics, structured illumination microscopy, or LIDAR applications. In this respect, exciting new opportunities are offered by all-optically tunable metasurfaces[19-25], which exploit the near-field enhancement of their resonant building blocks to manipulate different ultrafast nonlinear and optoelectronic effects. However, designing spatially variant structures for ultrafast complex wavefront control is difficult and thus, most existing works are limited to intensity modulation effects in periodic, homogeneous structures. As notable exceptions, Vabishchevich et al. demonstrated picosecond all-optical diffraction switching[26]; Iyer et al. achieved sub-picosecond steering of ultrafast incoherent emission[27]; Shaltout et al. combined a passive metasurface with a frequency-comb source to realize laser beam steering with a continuously changing steering angle[28]; and Di Francescantonio et al. realized all-optical routing of telecom photons upconverted to the visible range[29]. However, ultrafast all-optical wavefront control with high spatial complexity as e.g. required for vortex beam switching has not been realized so far.

A new avenue for ultrafast all-optical switching is offered by the unique nonlinear optical properties of monolayers of transition metal dichalcogenides (TMDs)[30]. Their crystal symmetry, combined with their atomically thin nature which relaxes phase matching constraints, allows for ultrafast all-optical modulation of the second harmonic (SH) field[31], at switching speed only limited by the pulse duration. However, despite its far-reaching potential, the mechanism has up to now only been employed for polarization and amplitude modulation. Here we demonstrate, for the first time, ultrafast all-optical switching of complex wavefronts using a cascaded meta-optical system consisting of a monolayer TMD, a quarter-wave plate, and a silicon metasurface. Specifically, we perform spatially resolved pump-probe experiments to demonstrate second harmonic beam deflection, vortex on-off switching and topological charge switching at femtosecond time scales, thereby increasing the speed of OAM modulators by six orders of magnitude. Importantly, our approach allows for switching between any two arbitrary wavefronts with pulse-duration limited dynamics. Our results can thus enable new technological developments in the fields of high-speed communications, remote sensing, ultrafast optics and holographic techniques.

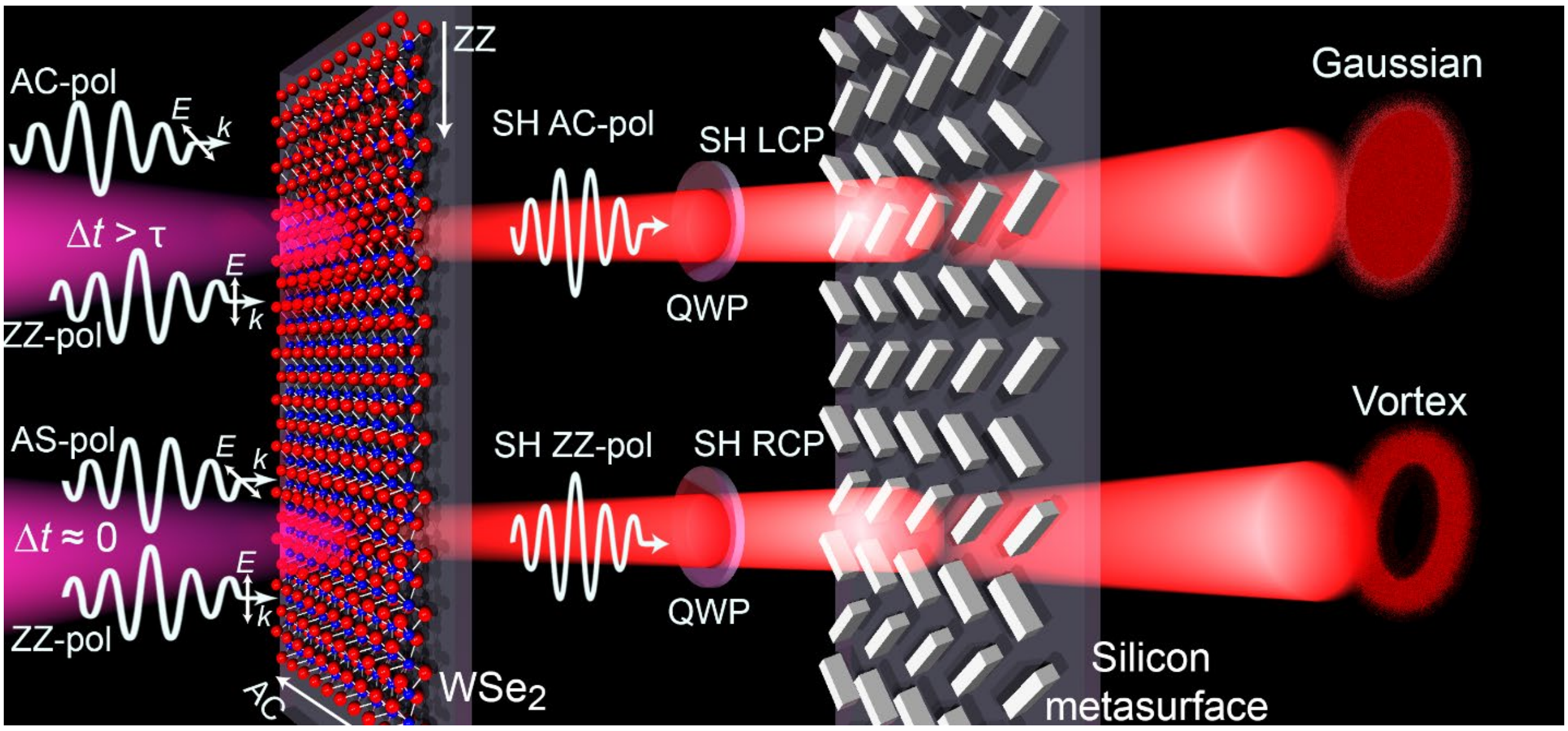

**Figure 1 | Experimental scheme of our suggested ultrafast wavefront shaping scheme for the example of Gaussian-to-vortex beam switching.**

A sketch illustrating the proposed ultrafast all-optical wavefront switching scheme is presented in Fig. 1. We use a $WSe_2$ monolayer to generate SH, where the (linear) SH polarization is controlled by the delay between two synchronized femtosecond laser beams with orthogonal linear polarization oriented along the armchair (AC) and zigzag (ZZ) directions of the crystal. $WSe_2$ was selected due to the low energy of its excitonic resonances as compared to other materials of the 2D TMD family, allowing for resonantly enhanced SH emission near 750 nm and thus facilitating metasurface fabrication. For delay values $\Delta t$ exceeding the pulse duration $\tau$, the linear polarization of the SH is emitted along the AC direction, while for simultaneous arrival of the two pulses ($\Delta t=0$), the SH is emitted along the ZZ direction[31]. Using a quarter-wave plate, the linear AC (ZZ) polarization of the SH beam is converted into left (right) handed circular polarization. Finally, the circularly polarized SH light is transmitted through a polarization sensitive wavefront-shaping metasurface and the output is recorded using a CCD camera. Specifically, we target three different effects of ultrafast wavefront shaping, namely beam deflection, Gaussian-to-vortex beam switching and topological charge switching.

In order to experimentally implement the described scheme, in a first step a $WSe_2$ monolayer was produced by mechanical exfoliation[32]. A corresponding optical microscope image is shown in Fig. 2a. To confirm that the exfoliated crystal is indeed a monolayer as well as the absence of notable strain in the crystal, we further conducted photoluminescence spectroscopy and pump-polarization dependent SHG measurements. Figure 2 b,c summarize the polarization resolved SH spectra of the monolayer for two different delay times $\Delta t$. As expected, for $\Delta t > \tau$ (Fig. 2 b), the generated SH is linearly polarized along the AC direction, while for $\Delta t \approx 0$ (Fig. 2 c), it is linearly polarized along the ZZ direction. To shape the wavefront in a polarization sensitive fashion, metasurfaces composed of hydrogenated amorphous silicon nanoresonators were designed based on the principle of Pancharatnam-Berry phase[33]. For the metasurface supporting Gaussian-to-vortex beam switching, a spatially variant propagation phase[34] was additionally implemented in the design, as it requires an additional degree of freedom. This is the case since the local phases for the two different polarizations have no simple relation as for the other considered phase profiles. All three types of metasurfaces were fabricated on a glass substrate using electron-beam lithography and inductively coupled plasma etching using an established process[35]. For each of the three targeted wavefront shaping effects, a separate metasurface with its own unique geometry and corresponding spatially variant phase profiles was fabricated. Figure 2 d-f shows the target phase profiles for both circular polarizations, light microscope images and scanning electron micrographs of the fabricated metasurfaces. The nanoresonators of all metasurfaces have an elliptical cross section with a height of 475 nm..

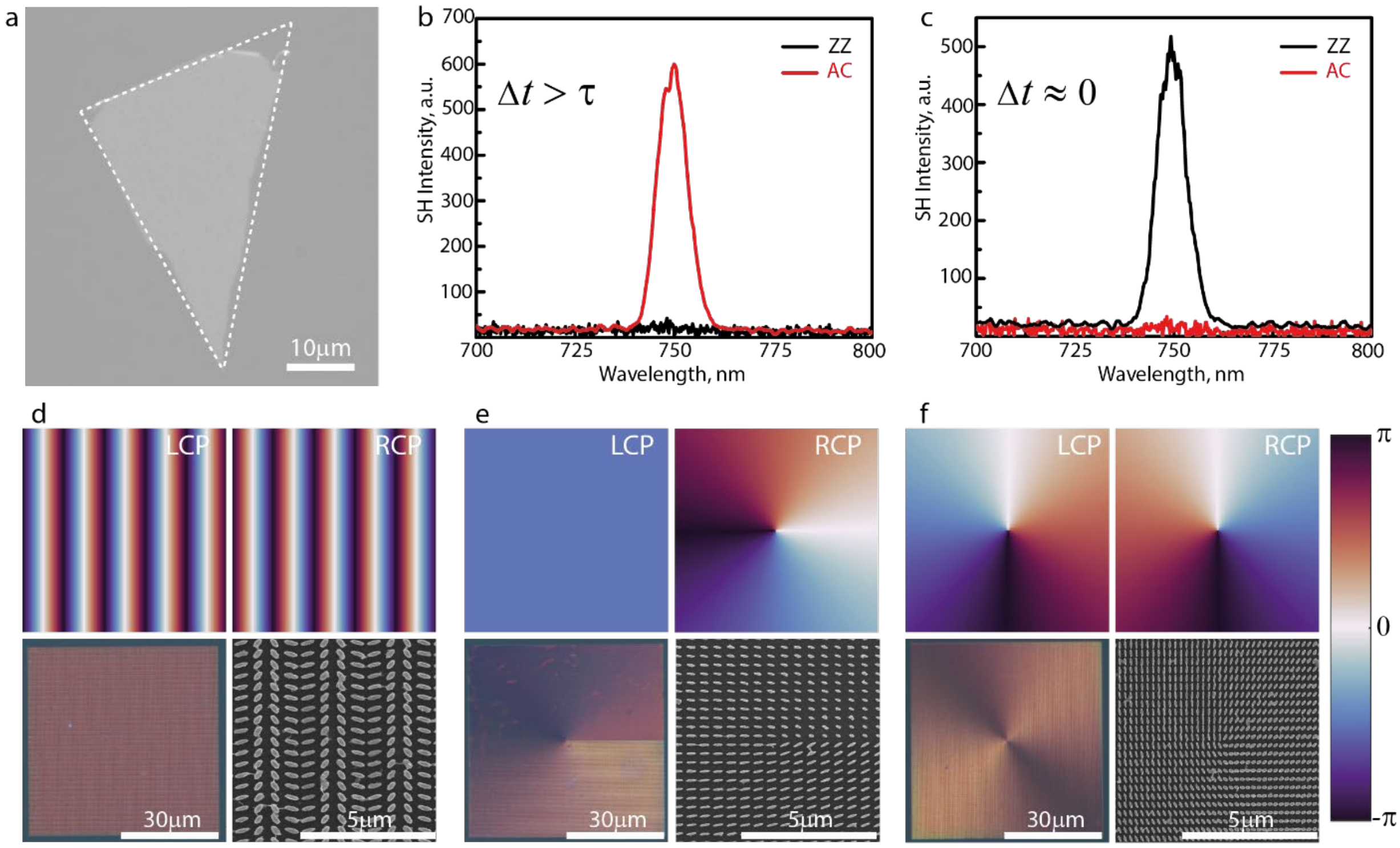

**Figure 2 | Components of the fabricated hybrid meta-optical system. a.** Optical microscope image of the employed $WSe_2$ monolayer. The white dotted line highlights the boundaries. **b-c.** SH spectra of the $WSe_2$ monolayer for $\Delta t > \tau$ **b** and $\Delta t \approx 0$ **c**. **d-f.** The top row shows the phase distribution for different circular polarizations of the incident field. The bottom row shows optical images of the metasurfaces (left) and SEM images of the central areas of the structures (right) for **d** beam deflection, **e** Gaussian-to vortex beam switching, and **f** topological charge switching, respectively.

Next, we numerically simulated the far-field distributions in Fourier space of a right-handed (RCP) and left-handed (LCP) circularly polarized Gaussian beam transmitted through the polarization-dependent metasurfaces using the commercial software package FDTD Lumerical (Fig. 3 a-f). The metasurface designed for ultrafast beam deflection implements the spatial phase profile of a blazed grating with the sign of the blaze reversing with handedness of the illumination polarization. Accordingly, the incident beam is dominantly channeled into the first diffraction orders, where the sign of the diffraction order changes with handedness (Fig. 3 a,c). For the metasurface sculpted for Gaussian-to-vortex beam switching, the spatial phase distribution for RCP light is flat. For LCP light, an azimuthal phase gradient ranging from 0 to $2\pi$ is implemented, resulting in the formation of a vortex beam with topological charge $l=1$. As expected, the intensity distribution behind the metasurface changes from a Gaussian distribution to a donut shape with a dark zone at the center caused by the topological singularity at the beam axis[1] (Fig. 3 b,e). Finally, the third metasurface is designed to implement an azimuthal phase gradient ranging from 0 to $2\pi$ for both incident polarizations, where the direction of the gradient reverses with handedness (36), corresponding to the formation of vortex beams with $l\pm1$, respectively. Note that for topological charge switching simple observation of the far-field distributions is not sufficient, since vortex beams with $l\pm1$ have identical far-field intensity patterns. Thus, to evaluate the sign of l of the output beam, we employ an astigmatic transformation, through which the intensity pattern of the vortex beams acquires the form of dark stripes, whose orientation changes with the sign of $l$ (Fig. 3 c,f). Experimentally, this transformation can be implemented by observing the intensity distribution in the focal plane of a cylindrical lens[37].

For an experimental demonstration of the various ultrafast wavefront shaping effects, we constructed a dedicated pump-probe setup featuring a Fourier microscope to observe the ultrafast change in the wavefront (see Methods for a sketch). As light source, we used the signal beam of an optical parametric oscillator (OPO) laser system (Mira OPO, APE, Germany) pumped by a femtosecond Ti:Sa laser (Coherent Chameleon Ultra II). The output wavelength was set to 1500 nm, the pulse repetition rate was 80 MHz, the pulse duration was 200 fs, and the power was 10 mW. We used a commercial common-path birefringent interferometer (GEMINI, NIREOS) to create two beams with orthogonal polarization and to control the time delay between them with attosecond precision and subwavelength interferometric stability. The superimposed beams were focused onto the $WSe_2$ monolayer using a Mitutoyo NIR M Plan 10x NA=0.26 objective, and the SH signal was collected using an identical objective. A short pass filter was used to filter out the beam at the fundamental harmonic frequency. Next, to convert the linear polarization of the SH beam (AC or ZZ) into circular polarization (LCP or RCP, respectively), a quarter-wave plate was employed. It should be noted that the orientation of the quarter-wave plate was fixed during the experiments, the polarization was changed only by the ultrafast nonlinear polarization switching effect in the $WSe_2$ monolayer as a function of the pulse delay. The circularly polarized SH beam was focused onto the metasurface using a Mitutoyo NIR M Plan 10x NA=0.26 objective, the transmitted signal was collected with a 20x NA=0.4 objective from the same manufacturer. Finally, a system of lenses was used to image the back focal plane of the collection objective onto an sCMOS camera (prime 95B, Teledyne, USA). For the case of topological charge switching, one of the lenses was replaced by a cylindrical lens. This lens introduces a controlled astigmatism, leading to different intensity distributions on the camera, depending on the sign of $l$. Figure 3 g-m show the recorded intensity distributions in the far field for different delay values $\Delta t$ corresponding to subsequent ($\Delta t > \tau$, left) and simultaneous ($\Delta t \approx 0$, right) arrival of the two pulses. For all three studied metasurfaces, the targeted ultrafast switching effects can be observed. As a manifestation of ultrafast beam deflection, Fig. 3 g,k shows the ultrafast change in the intensity channeled into the first diffraction orders of different sign. Figure 3 h,l shows the change from a Gaussian to a donut shaped intensity distribution as the temporal pulse overlap is established, thereby providing clear evidence of ultrafast Gaussian-to-vortex beam switching. Finally, Fig. 3 j,m shows intensity distributions characteristic for switching of the topological charge from $l$=-1 to $l$=+1 by changing the pulse delay. Deviations of the measured intensity distributions from the theoretical expectations can be explained by fabrication imperfections of the metasurface sample. For example, for beam deflection, a larger amount of light is also channeled into the zeroth diffraction order for both displayed settings of the delay stage.

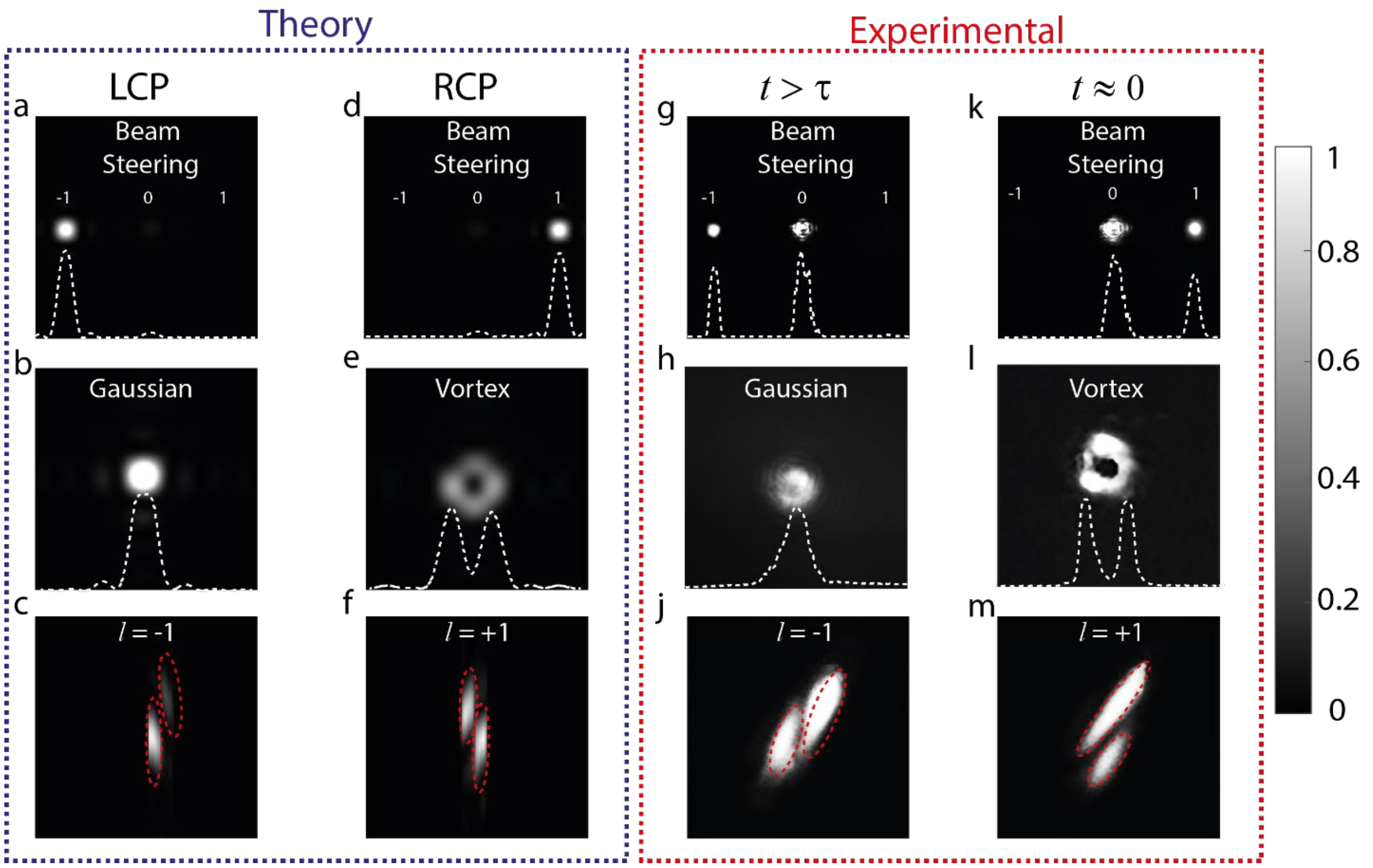

**Figure 3 | Theoretical (left) and experimental (right) far-field distributions**. **a-c**, **g-j** LCP and **d-f**, **k-m** RCP light transmitted through the polarization-dependent metasurfaces in Fourier space. The first row shows results for beam deflection, the middle row for Gaussian-to-vortex beam switching, and the bottom row for topological charge switching.

Finally, in order to directly monitor the ultrafast dynamics of the observed complex wavefront shaping effects, we recorded the far-field intensity as a function of the delay time Δ*t* for the case of Gaussian-to-vortex beam shaping. To ease the representation, instead of the full two-dimensional image, a cross section through the center of the beam is displayed. Note that for overlapping pulses, the polarization of the GEMINI output changes systematically between linear, elliptical and circular polarization as the delay time is changed by less than an optical cycle. Thus, from the full interferometric trace we extracted only those frames which correspond to a linear output polarization (Fig. 4). Clearly, if the delay is less than the pulse duration (i.e. Δ*t*<~100 fs), we observe the typical cross section of a vortex beam, namely two maxima and a central minimum of intensity. In contrast, when the delay exceeds the pulse duration, the cross section is characterized by a single intensity maximum typical of a Gaussian beam.

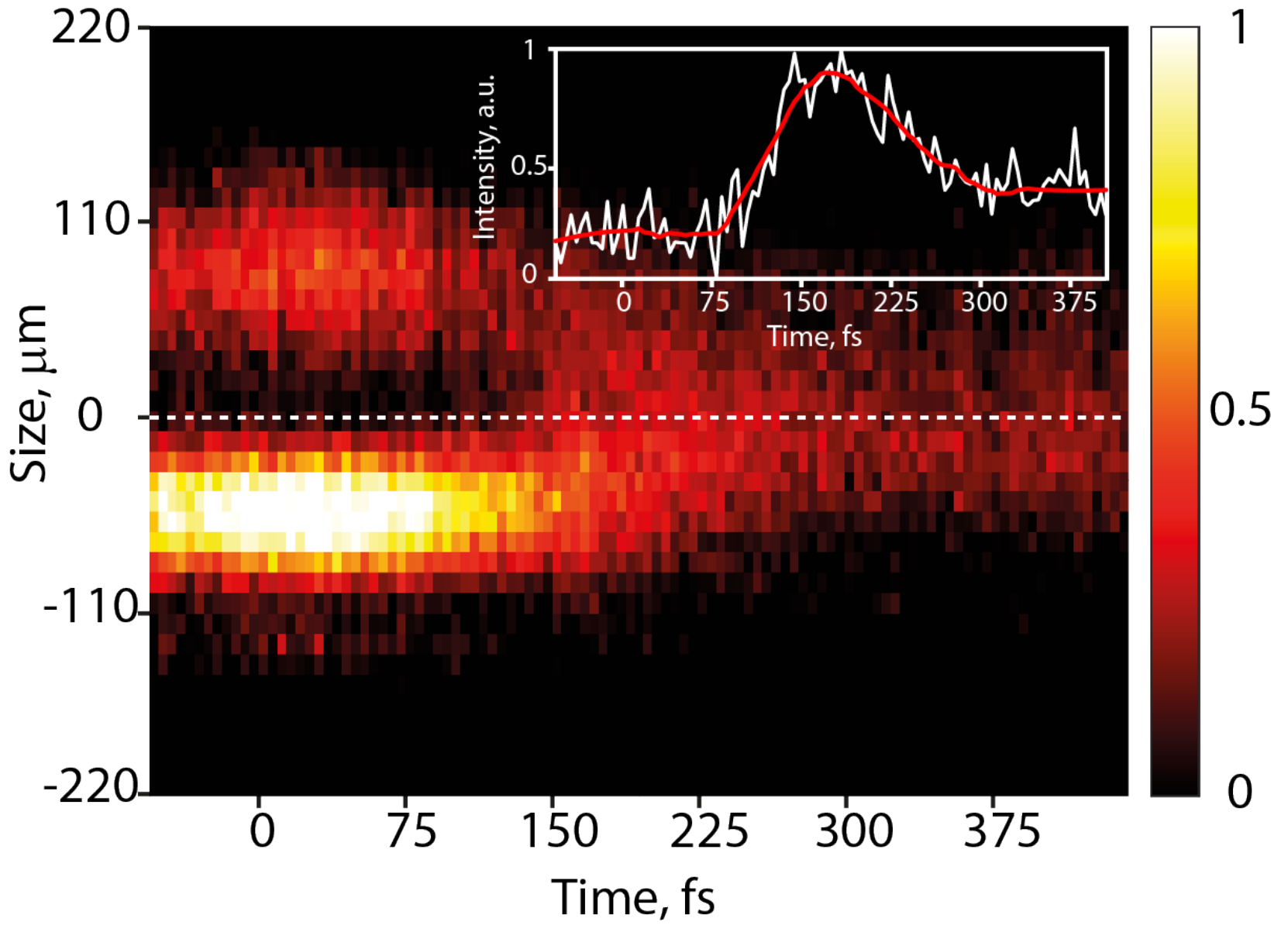

**Figure 4| Ultrafast wavefront shaping dynamics.** The map shows the cross section of the far-field intensity as a function of the delay time Δt for the case of Gaussian-to-vortex beam switching. The inset shows the intensity in the center of the map (dashed white line) versus time. The red curve represents a guide to the eye obtained by smoothing the experimental data using a Savitzky–Golay filter.

In conclusion, we experimentally demonstrated femtosecond all-optical wavefront control using polarization-dependent wavefront-shaping metasurfaces. This was accomplished by combining all-optical second-harmonic polarization switching from a monolayer TMD with a polarization-sensitive wavefront shaping metasurface. We demonstrated three different wavefront shaping effects, namely beam deflection, Gaussian-to-vortex beam switching and topological charge switching, with response times in the femtosecond range. Our approach allows for switching between any two arbitrary wavefronts with pulse-duration limited dynamics and thus opens a pathway towards ultrafast OAM encoding of information for free space optical communication links. More generally, the demonstrated speed-up in the synthesis of complex arbitrary wavefronts by more than 6 orders of magnitude offers wide-reaching opportunities for both fundamental science and applications in high-speed communications, remote sensing, ultrafast optics and holographic techniques. Importantly, since the underlying SHG characteristics are a consequence of the crystal structure of the $WSe_2$ monolayer and thus broadband, the demonstrated scheme can be scaled to other wavelengths – albeit with varying SH generation efficiency – by simple adjustment of the metasurface geometry. Also, an increase of SH output power could be achieved within the limits dictated by phase-matching considerations by replacing the $WSe_2$ monolayer by a bulk 3R-$WSe_2$ (rhombohedral) crystal.

### Data availability

Data that support the plots within this paper and other findings of this study are available from the corresponding author upon reasonable request.

## Methods

### Numerical simulations

The simulations of the far field of the metasurfaces were performed using the commercial software package FDTD Lumerical. Each metasurface consists of 20 by 20 nanoposts implementing the required phase profile and resting on a glass substrate. The truncation of the metasurface allows fast computation, while retaining a reasonable representation of the physical system. In the simulations, a circularly polarized pulse with plane wave profile excites the metasurface in a domain enclosed by perfectly matched layer (PML) boundaries. The output field is recorded and Fourier-transformed to provide the spectral static field distribution. The far-field patterns of the metasurfaces were projected from the near field at the plane immediately after the nanoposts.

To evaluate the topological charge of the output of the metasurface implementing topological charge switching, a one-dimensional Fourier transform was performed on their corresponding far-field pattern to imitate the effect of a cylindrical lens.

### Fabrication of the $WSe_2$ monolayer

Tungsten diselenide ($WSe_2$) monolayers were mechanically exfoliated from a commercially available bulk crystal (HQGraphene). The monolayers were subsequently transferred onto a fused silica substrate by a deterministic dry-transfer technique using a polydimethylsiloxane (PDMS, GelPak PF-40-X4) stamp. The monolayers were pre-characterized in a custom-built optical microscope by photoluminescence spectroscopy as well as optical contrast measurements.

## Acknowledgements

We acknowledge Marijn Rikers for help with SEM images. A.D.S. acknowledges Dr. Ivan Shishkin for the discussion on optical measurements. This project has received funding from the European Union's Horizon 2020 research and innovation programme under the H2020-FETOPEN-2018-2020 grant agreement no. [899673] (Metafast); TP, IS, GS, GDV. This work was also partially funded by the Deutsche Forschungsgemeinschaft (DFG, German Research Foundation) through the Collaborative Research Centre SFB 1375 ("NOA", projects B2, B5 and C5), the International Research Training Group (IRTG) 2675 "Meta-ACTIVE", project number 437527638, and through the Emmy Noether Program, project number STA 1426/2-1.

## Author contributions

A.S., S.H.L., F.C., S.K. performed the optical measurements. D.Y.C., S.K. fabricated the meta-optical system. S.H.L., F.C. performed theoretical calculation. A.S., I.S. wrote the first draft of the manuscript. I.S., T.P., G.S., G.D.V. funding acquisition. I.S., G.S., G.D.V. proposed the idea. I.S., T.P., G.S., G.D.V. supervision and project administration. All authors analyzed the results and contributed to the writing and discussion of the manuscript.

# Supplementary Information

## Design principle of metasurface

Three metasurfaces were designed to realize different ultrafast-switching functions, namely Gaussian-to-vortex-beam switching, topological charge switching and beam steering. The first metasurface switches between vortex beam and non-vortex beam, the second switches between vortex beams with +1 and -1 topological charge, while the last metasurface switches the beam steering from a positive angle to a negative angle. The required phase profiles, $\phi_{\mathrm{LCP}}(x,y)$ and $\phi_{\mathrm{RCP}}(x,y)$, in the two operating modes of the three metasurfaces respectively are summarized in Table S1. The modulation of phase of light by the particles on the metasurface is based on the principle of Pancharatnam-Berry phase, to be explained in this section. For the metasurface for vortex beam on/off switching, in which an addition degree of freedom for phase modulation is required, the manipulation of propagation phase was also employed in the design.

Consider an elliptic-cylindrical meta-atom with its long- and short- axis aligned with the *x*- and *y*-axis, the Jones matrix of the meta-atom reads

$$\widehat{T_0} = \begin{pmatrix} T_{xx} & 0 \\ 0 & T_{yy} \end{pmatrix}.$$

A rotation of the meta-atom along the $z$-axis with an angle θ would result in the modified Jones matrix of the meta-atom:

$$\hat{T}(\theta) = \hat{R}(-\theta)\widehat{T_0}\hat{R}(\theta) = \begin{pmatrix} T_{xx}\cos^2\theta + T_{yy}\sin^2\theta & (T_{xx}-T_{yy})\cos\theta\sin\theta \\ (T_{xx}-T_{yy})\cos\theta\sin\theta & T_{xx}\sin^2\theta + T_{yy}\cos^2\theta \end{pmatrix},$$

where $\hat{R}(\theta) = \begin{pmatrix} \cos\theta & \sin\theta \\ -\sin\theta & \cos\theta \end{pmatrix}$ is the rotation matrix. An input LCP/RCP light $E_{\pm} = (1 \pm i)$ would result in an output light with two parts:

$$E_{\pm}^{\mathrm{out}} = \hat{t}(\theta)E_{\pm} = 1/2(T_{xx}+T_{yy})E_{\pm} + 1/2(T_{xx}-T_{yy})e^{\pm i2\theta}E_{\mp},$$

where the subscript in $E_{\pm}^{\mathrm{out}}$ indicates the input polarization.

The two parts are the co-polarized output, of which the polarization is the same as the input, and the cross-polarized output, of which the polarization is orthogonal to the input. The phase of the cross-polarized output scales linearly with the geometric rotation angle θ of the meta-atom, while the direction of scaling depends on the handedness of the input light. Employing properly designed meta-atom which suppress the co-polarized output would allow the manipulation of phase of output light through the geometric phase 2θ.

The geometric phase provides one degree of freedom for the spatial control of phase as the input light propagates through the metasurface. This one degree of freedom is sufficient for constructing a polarization-dependent metasurface in which the phase modulation profile $\phi(x,y)$ of one polarization is the negative of the other, i.e. $\phi_{\mathrm{LCP}}(x,y) = -\phi_{\mathrm{RCP}}(x,y)$. The vortex beam charge switching and the beam steering metasurface can hence be solely based on the principle of geometric phase. The vortex beam on/off metasurface however requires a spatial profile different from that allowed by merely geometric phase. In such cases, the propagation phase which light acquires during the propagation in the meta-atom is also exploited. The geometric phase controlled by the rotation angle of the meta-atom and the propagation phase controlled by the dimension of the meta-atom provide two degrees of freedom and allow an arbitrary phase profile for each polarization input.

The design of the metasurfaces required the search for a suitable set of meta-atoms which suppresses the co-polarized output and allows 2π-propagation phase modulation. The conditions can be written in terms of the Jones matrix components as follow:

1. $|T_{co}|^2 = |1/2(T_{xx} + T_{yy})|^2$ is minimized,
2. $\arg(T_{cross}) = \arg(1/2(T_{xx} - T_{yy}))$ spans the range $[-\pi, \pi)$.

For the need to operate in the non-resonant regime that maximizes transmission, elliptical nanopost was selected to be the meta-atom. Hydrogenated amorphous silicon (a-Si:H) was appointed as the meta-atom material for its availability of fabrication resources and low absorption at the operation wavelength (750 nm). The lattice constant $p$ of the meta-atoms array was set at 450 nm such that the metasurface is in the non-diffractive regime, while the inter-particle interaction is negligible.

The Jones matrix components $T_{xx}$ and $T_{yy}$ of a-Si:H nanoposts with different height $h$, semi-major and minor axes, $r_x$ and $r_y$ at wavelength 750 nm were evaluated using the finite difference method in frequency domain with CST Studio Suite. The system was modelled as periodic identical nanoposts on glass substrate with lattice constant $p$. The model of nanopost was built with a side wall angle $\alpha$ of 3.5° to resemble the fabrication limit. The refractive index of nanoposts for the simulation were experimentally measured from a sample of a-Si:H ($n$ = 3.77), while the refractive index of the substrate is set to 1.45. A schematic of the nanoposts is provided in Fig. S1.

From the calculation result, it is observed that with height $h$ = 490 nm it is possible to provide a set of nanoposts, which fulfill our optimization condition. Among all nanoposts with $h$ = 490 nm, the nanoposts with $r_x$= 45 nm shows minimal co-polarized output, while $\mathrm{Arg}(T_{\mathrm{cross}})$ spans approximately full 2π range by varying $r_y$ (Fig.S2). Based on the above result, the height of the nanoposts for all metasurfaces were set to 490 nm. Nanoposts with $r_x$ = 45 nm and $r_y$ ranging from 60 nm to 180 nm were included in the vortex beam on/off switching metasurface to fulfill the 2π variation of propagation phase. The vortex beam charge switching metasurface and beam steering metasurface, which do not require the manipulation of the propagation phase of the nanoposts, were constructed with only one type of a-Si:H nanopost ($r_x$ = 60 nm and $r_y$ = 185 nm), which gives maximal contrast of cross-polarized output to co-polarized output. The models of the three metasurfaces constructed with the nanoposts according to the corresponding phase profiles are shown in Fig. S3.

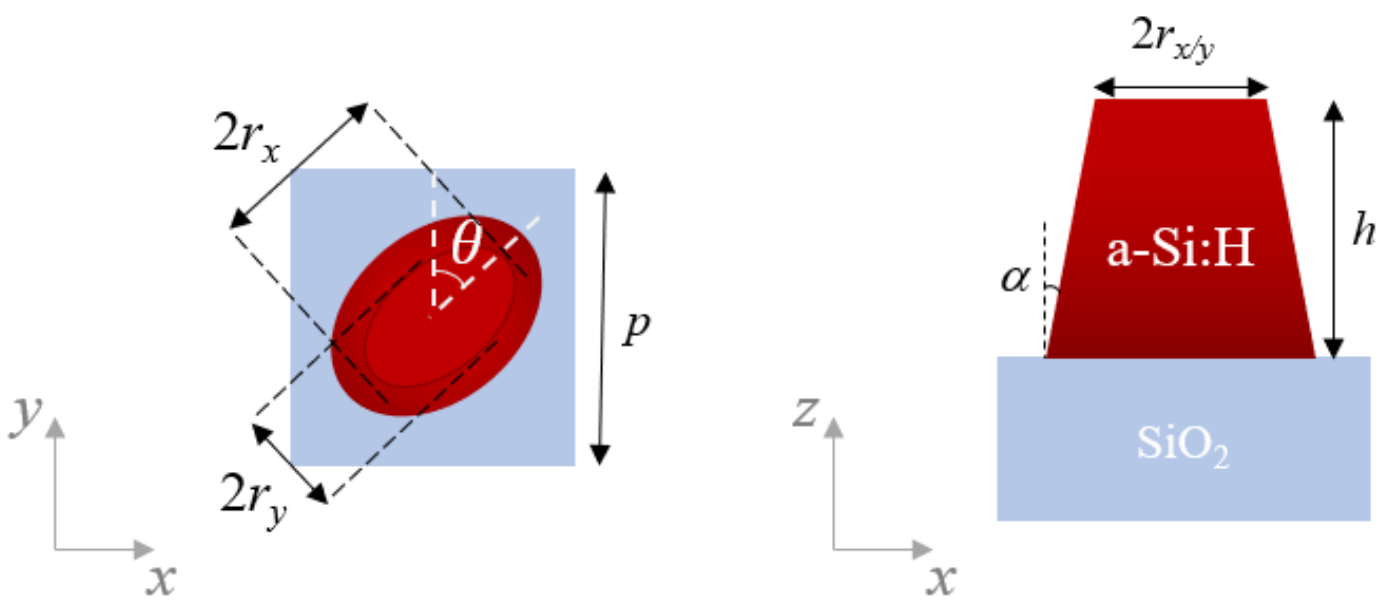

Figure S1. Schematics of a nanopost lattice. The lattice constant $p$, semi-major and minor axes $r_x$ and $r_y$, rotation angle $\boldsymbol{\theta}$ and height $h$ of the nanopost, and the Cartesian coordinates system are specified.

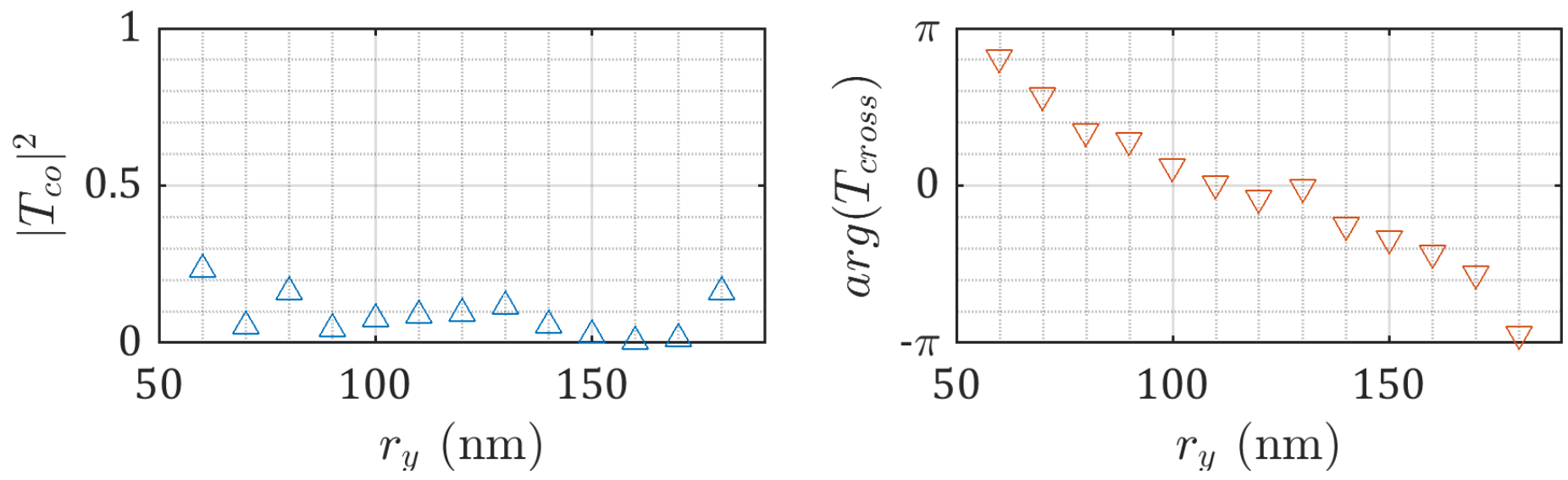

Figure S2. Transmission $|\boldsymbol{T}_{\mathrm{co}}|^2$ and phase Arg($\boldsymbol{T}_{\mathrm{cross}}$) of nanoposts employed in vortex beam on/off metasurface. The height $h$ and semi-major axis $r_x$ of the nanoposts are 490 nm and 45 nm, respectively.

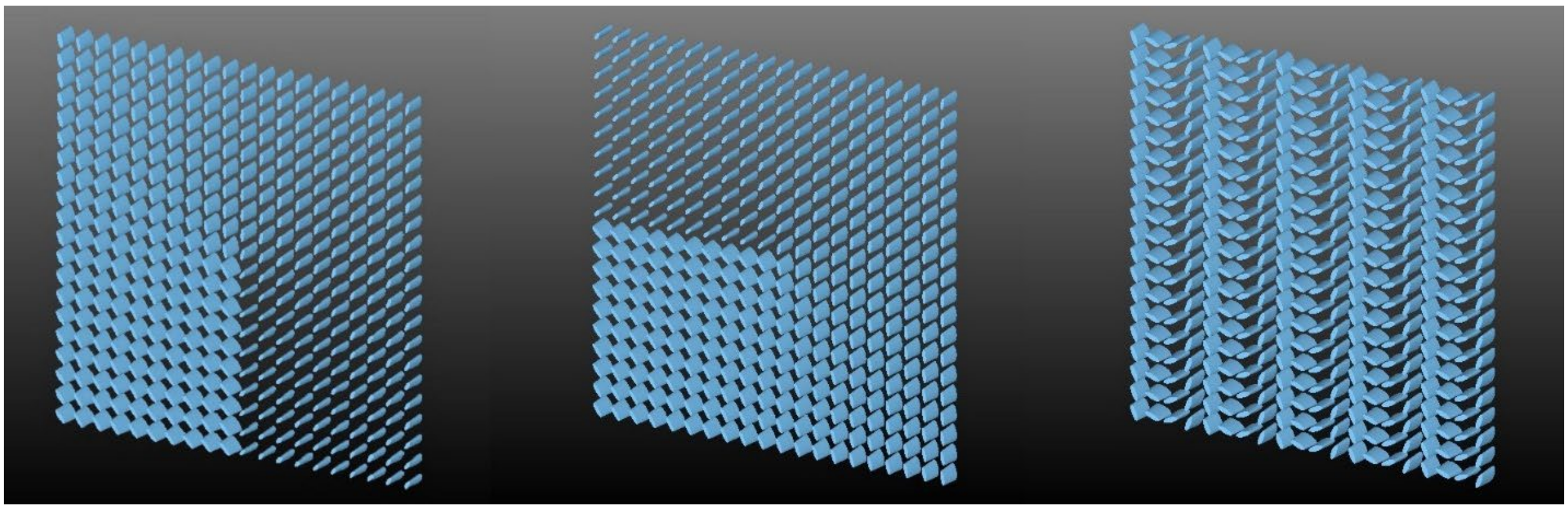

Figure S3. 3D models of the metasurfaces for vortex beam on/off switching (left), vortex beam charge switching (center) and beam steering (right).

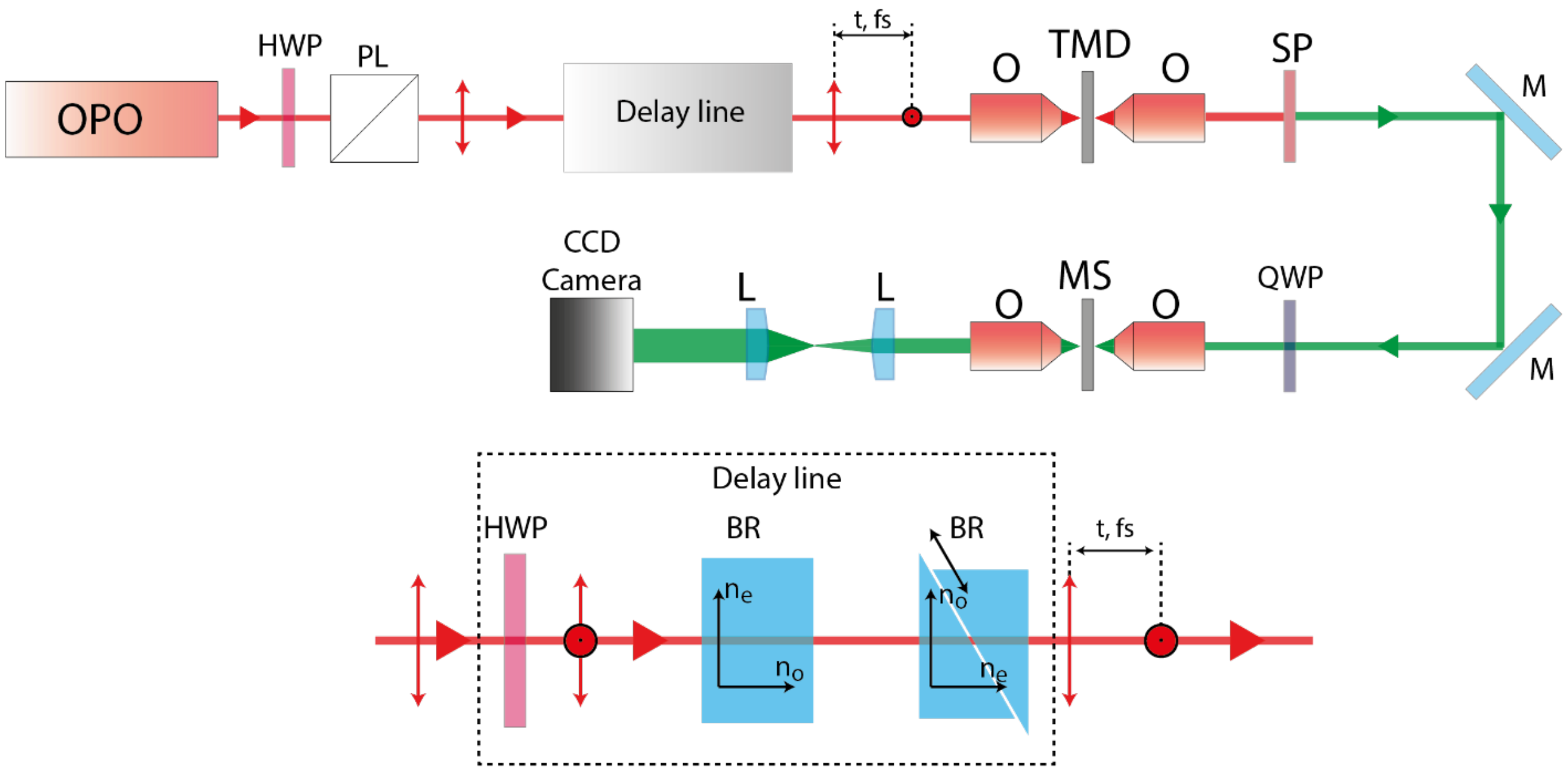

Figure S4. Optical scheme of the home-built setup used for ultrafast second-harmonic wavefront control

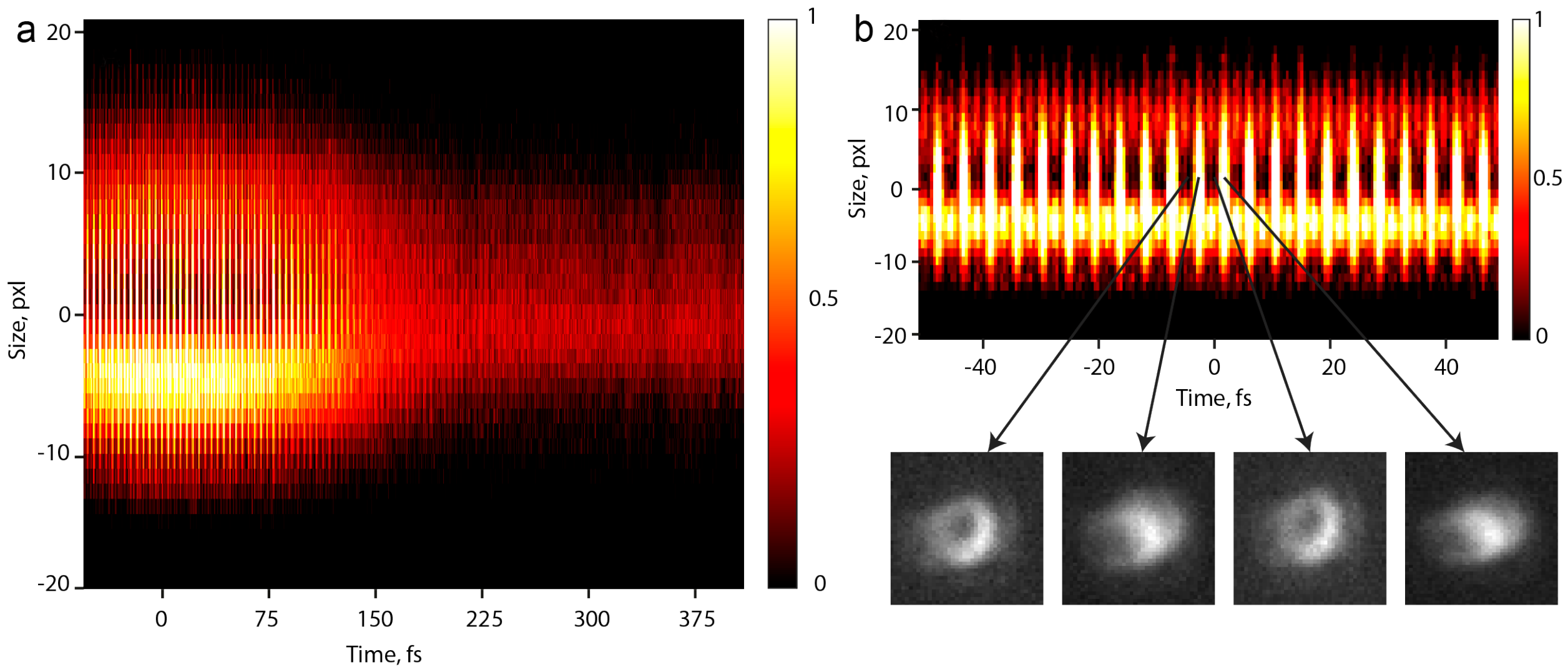

Figure S5 Full interferometric trace. **a.** The map shows the cross section of the far-field intensity as a function of the delay time $\Delta t$ for the case of Gaussian-to-vortex beam switching. Note that in contrast to Fig. 4 in the main manuscript, here the full interferometric trace is shown, including data corresponding to GEMINI interferometer outputs with elliptical and circular polarization. **b.** Fragment of **a** showing a close-up of the interference fringes. The insets show the spatial SH intensity distribution at the marked positions. Doughnut shapes are clearly observed for delays corresponding to linear output states of the interferometer.

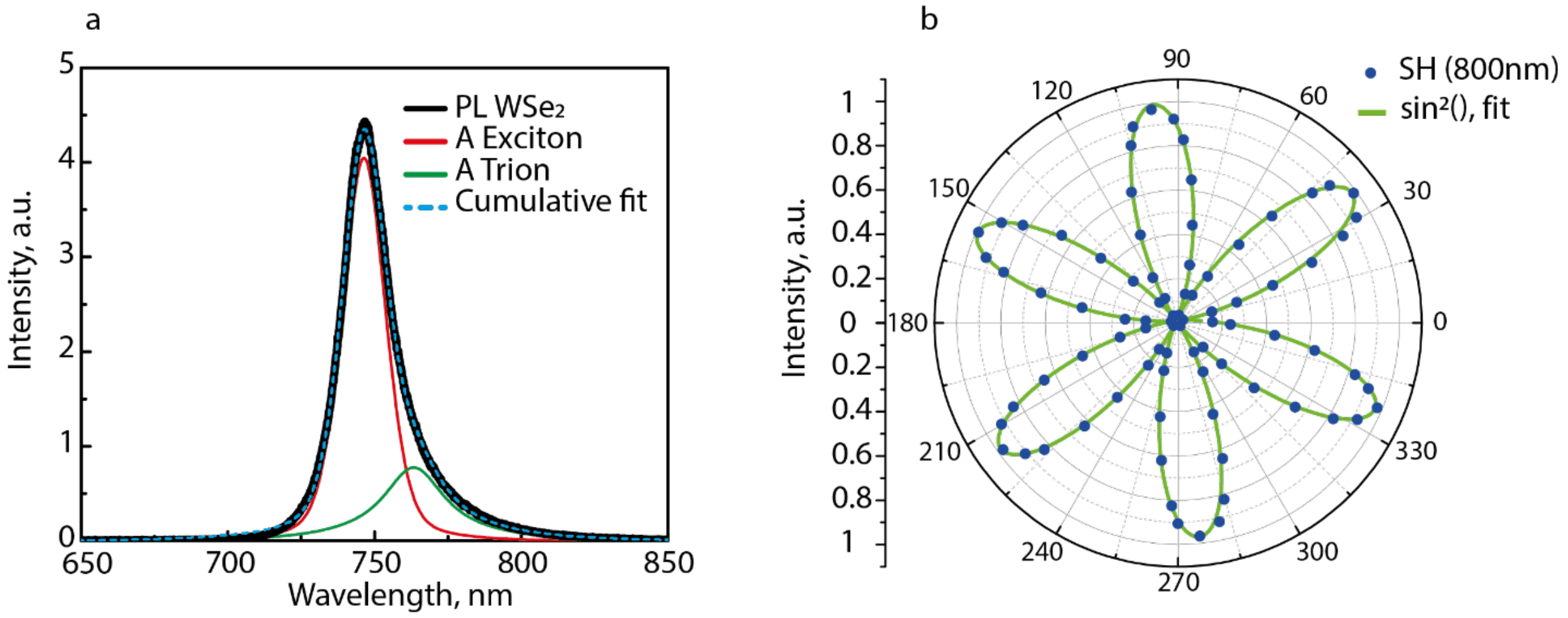

Figure S6 Optical characterization of the $WSe_2$ monolayer. **a.** Photoluminescence (PL) of the $WSe_2$ monolayer. For the excitation we used a 532 nm CW-laser (Cobolt 08-DPL 532 nm), which was attenuated to an average power of 100 μW and focused onto the sample with an x50 microscope objective (M Plan Apo HL; NA: 0.42). The backscattered PL (black) was subsequently detected with a spectrometer (Horiba, iHR 550 with a Synapse PLUS CCD). Pseudo-Voigt fits show the contributions for exciton (red) and trion (green), respectively. The cumulative fit (dashed blue) agrees very well with the experimental data. **b**. Polar plot of the normalized SH intensity a function of the excitation polarization angle δ. The detected SH polarization was always selected parallel to the excitation with a co-rotating wiregrid polarizer (Thorlabs, WP25M-UB) in transmission geometry[1]. As excitation we used the signal output of an OPO (Levante IR from APE) at 1600 nm (10 mW average power; 150 fs pulse duration; 76 MHz repetition rate) which was focused on the monolayer with a x40 reflective objective (Thorlabs, LMM40X-UVV). The blue dots show experimental data and the solid green line indicates the $\sin^2[3(\delta - \delta_0)]$ fit. From the perfectly symmetric six-fold pattern we conclude that no notable strain is present in the investigated sample.

| METASURFACE FUNCTIONS | $\phi_{\mathrm{LCP}}(x,y)$ | $\phi_{\mathrm{RCP}}(x,y)$ |
|---|---|---|
| VORTEX BEAM ON/OFF | 0 | $+\mathrm{Arg}(\boldsymbol{x}+\boldsymbol{iy})/\mathbf{2}$ |
| VORTEX BEAM CHARGE | $+\mathrm{Arg}(\boldsymbol{x}+\boldsymbol{iy})/\mathbf{2}$ | $-\mathrm{Arg}(\boldsymbol{x}+\boldsymbol{iy})/\mathbf{2}$ |
| BEAM STEERING | $+\boldsymbol{ax}$ | $-\boldsymbol{ax}$ |

Table S1. A summary of the phase profiles in LCP and RCP modes for metasurfaces with different functions, respectively.